# Spin-wave dynamics in Permalloy/Cobalt magnonic crystals in the presence of a non-magnetic spacer

P. Malagò[1], L. Giovannini[1], R. Zivieri[1], P. Gruszecki[2], M. Krawczyk[2]

[1] Dipartimento di Fisica e Scienze della Terra, Università di Ferrara, Via G. Saragat 1, I-44122 Ferrara, Italy
[2] Faculty of Physics, Adam Mickiewicz University in Poznań, Umultowska 85, 61-641 Poznan, Poland

Abstract

In this paper, we theoretically study the influence of a non-magnetic spacer between ferromagnetic dots and ferromagnetic matrix on the frequency dispersion of the spin wave excitations in two-dimensional bi-component magnonic crystals. By means of the dynamical matrix method we investigate structures inhomogeneous across the thickness represented by square arrays of Cobalt or Permalloy dots in a Permalloy matrix. We show that the introduction of a non-magnetic spacer significantly modifies the total internal magnetic field especially at the edges of the grooves and dots. This permits the manipulation of the magnonic band structure of spin waves localized either at the edges of the dots or in matrix material at the edges of grooves. According to the micromagnetic simulations two types of end modes were found. The corresponding frequencies are significantly influenced by the end modes localization region. We also show that, with the use of a single ferromagnetic material, it is possible to design a magnonic crystal preserving properties of bi-component magnonic crystals and magnonic antidot lattices. Finally, the influence of the non-magnetic spacers on the technologically relevant parameters like group velocity and magnonic band width are discussed.

## I. Introduction

Spatial periodicity in a ferromagnetic material modifies spin wave (SW) dispersion relation and results in the formation of magnonic bands and band gaps. Magnetic materials with periodic modulation are called magnonic crystals (MCs) [1,2,3]. Presently, MCs get particular interest due to the possibility of tailoring frequency spectra of SWs at the nanoscale; as a consequence, it is possible to understand magnetization dynamics and: a) to design metamaterial devices[4,5], b) to transduce and transmit signals[6,7,8], c) to realize magnonic transistors[9], d) to make logic operations[10,11,12].

Among the possible geometries of MCs, the planar MCs are the most often investigated. This is due to the feasibility of fabrication of regular patterns and easy access to characterize magnetic properties, to measure SW dispersion relation and dynamics in time domain and to visualize SW excitations[13,14]. In standard SW transmission measurements microwave transducers (microstripes or coplanar waveguides) are used. They allow for effective excitation of SWs with long wavelengths. In this limit the magnetostatic interactions are important and propagation of SWs in nanostructures is usually investigated in the direction perpendicular to the external magnetic field, i.e., in the Damon-Eshbach (DE) geometry, where even at zero wavenumber the relatively high group velocity is present.

Among planar MCs, the one- and two-dimensional (2D) MCs, i.e., with periodicity along one and two directions, respectively, can be distinguished. The three main groups of 2D MCs are: arrays of dots, antidot lattices and bi-component MCs (BMC). The first consists of regular arrays of thin ferromagnetic dots, the second of negative arrays of the former, i.e., arrays of holes in thin ferromagnetic film. The last group can be regarded as a superposition of both, i.e., the antidot lattices with holes filled with a different ferromagnetic material. These three groups present distinct features in the SW propagation. The collective magnetization dynamics in an array of dots is solely due to dynamic dipole coupling between resonant excitations of the dots, however its properties are also influenced by static demagnetizing effects[15]. In the case of weak coupling (large separation between dots with respect to their thickness and width), the magnonic spectra consist of flat bands with frequencies related to the eigenmode excitations of the isolated dot[16]. By increasing the dynamic dipole coupling, e.g., by decreasing separation between dots, collective SW excitations with finite band width and preserving properties of the magnetostatic waves appear. The widening of the bands depends on the dipolar coupling strength and on the stray magnetic field[17]. However, the dynamic dipole interaction is effective especially for eigenmodes having the largest total dynamic magnetization (averaged over the whole dot), viz. mainly for the fundamental mode[18], but also for end modes or low-order backward-like magnetostatic modes[19]. In antidot lattices, the low frequency part of SW spectra is influenced by the inhomogeneous static demagnetizing field created by the edges of the holes. The presence of holes leads to the formation of wells of the total magnetic field where magnetization dynamics mainly concentrate[20,21]. Indeed, in antidot lattices end modes localized at the edges of the holes and SWs concentrated in channels between holes were found[22,23]. These effects disappeared at sufficiently small lattice constants, where the exchange interactions starts to prevail over the dipole interactions. Recently, also the effect of magnetization pinning on spin-wave dispersion has been theoretically studied in Permalloy (Py) antidot waveguides by introducing a surface anisotropy at the ferromagnetic/air interface[24]. Moreover, it has been shown that structural changes in antidot waveguides breaking the mirror symmetry of the waveguide can close bandgaps[25].

It is also well known from the literature that the Dzyaloshinskii–Moriya interaction induces the tilting of the magnetic moments at the edges and leads to the formation of a non-collinear structure[26] acting as a scattering barrier for spin excitations[27] and partly contributes to the formation of end modes along the barrier. The transition from quantized to propagative regime of SWs (end modes and fundamental mode) can be controlled e.g., by the magnetic field orientation or by the separation between holes[28,29,30]. In addition to the demagnetizing effects, also the shape and size of holes in the antidot lattices influence the SW spectrum. This effect dominates for exchange SWs, i.e., at high frequencies or when the lattice constant is small[31]. In bi-component MCs the inhomogeneous demagnetizing field is still present, however its amplitude depends on the difference between the magnetic properties of the constituent materials. Thus, its influence on SW dynamics is weaker than in antidot lattices and valuable for the low-frequency modes only.

Recently, a bi-component MC composed of Co circular dots embedded in a Py ($Ni_{80}Fe_{20}$) matrix was investigated theoretically and experimentally[14,32,33,34]. The Brillouin light scattering (BLS) measurements showed the existence of two types of SW excitations concentrated in regions perpendicular to the external magnetic field containing Co dots and in Py matrix between the Co dots[32]. Theoretical studies have confirmed that the separation of frequencies of these SWs is due to a magnetostatic effect[33,34] and the splitting of the magnonic band at the boundary of the Brillouin zone (BZ) is connected to the periodicity of the magnetic system[14]. However, the full magnonic band gap in bi-component MCs has not yet been investigated in detail. Changes of dot or antidot shape, their rotation with respect to the crystallographic axes, imperfections in their shape or at their edges can further modify SW spectra[35,36,37]. Thus, the large variety of shapes for dots or antidots and of their arrangements together with magnetic configurations which can be realized in MCs[38,39,40,41], makes magnonics an inexhaustible and intriguing topic of research. More specifically, no enough attention has been given to the study of collective dynamics in bi-component MCs where a non-magnetic spacer separates the two magnetic materials. The aim of this paper is thus to theoretically investigate the effect of a non-magnetic spacer in 2D MCs on the dispersions of the relevant SWs according to a micromagnetic approach named Dynamical Matrix Method (DMM). This is done to investigate the important spin dynamics effects due the significant spatial variations experienced by the total inhomogeneous magnetic field because of the non-magnetic material at the interface between the two ferromagnetic materials. In this respect, it is studied the dynamics in square lattice 2D MCs with square antidots partially filled with different magnetic materials but the obtained results can be easily generalized to other geometries. This is achieved by putting the non-magnetic spacer around the dots embedded in antidot lattices. In this study two types of separation between dots and Py matrix are considered: a) with the non-magnetic spacer located only below the dot, and b) with the spacer fully around the dot. It is shown that these separations create an inhomogeneous static demagnetizing field which allows for the formation of end modes in the matrix (characteristic for the antidot lattices) and end modes in the dots (characteristic for the array of dots) which were not yet found in the previous studies[14,34]. Moreover, it is shown that similar properties can be achieved using a single ferromagnetic material, i.e., in single component 2D MC. This study focuses on the important part of magnetism devoted to SW phenomena in composite structures, which is almost unexplored yet in the case of large scale 2D bi-component nanopatterned systems. This investigation is also of interested for technological applications in the area of magnonics, magnetic memories and metamaterials.

The paper is organized as follows. In the next section the structures and the theoretical method used in the investigations are described. Then, in Sec. III, the results of calculation of SW spectra showing the influence of non-magnetic spacers on the magnonic dispersions are presented. In Sec. IV the results obtained are discussed and the influence of the inhomogeneity of the total magnetic field is analyzed. Then, in Sec. V, the effect of the non-magnetic spacer on the group velocity and magnonic band width is investigated. Finally, conclusions are drawn in Sec. IV.

## II. Structure and method

In order to study the dynamical properties of 2D MCs connected with the non-magnetic spacers between ferromagnetic materials, the dispersion relations of SWs for five systems have been calculated. The magnetic systems are composed of Py, Co and non-magnetic material. All geometries investigated here are based on square lattice and square magnetic dots, the lattice constant being $a = 400$ nm. MCs are supposed to be infinite in plane (along $x$ and $y$). These five systems are depicted in Fig. 1: (a) **System 1 (S1)**: bi-component MC composed of 30 nm thick Py film with an array of 20 nm deep square grooves of 200 nm size. In the bottom of grooves there is 10 nm of non-magnetic material and then Co dots (20 nm thick) partially immersed into the grooves. The Co dots are in direct contact with Py only at lateral edges of the dot. (b) **System 2-Co ($S2^{Co}$)**: bi-component MC similar to S1 but with 10 nm width spacer around the Co dots (200 nm wide). In $S2^{Co}$, Co dots and Py matrix are separated by a non-magnetic spacer. (c) **System 2-Py ($S2^{Py}$):** one component MC with the same geometry of $S2^{Co}$ but with Py dots. (d) MC composed of square Py dots (10 nm thick and 200 nm wide) surrounded by non-magnetic spacer and fully immersed in the Py matrix. This is **system 3 (S3)**. (e) An array of squared Co dots (20 nm thick and 200 nm wide) constitutes the **system 4 (S4)**. All parameters used in the simulations are typical parameters for Py and Co materials[42,43] : saturation magnetization for Py $M_{S,Py} = 750$ emu/cm$^3$ and for Co $M_{S,Co} = 1200$ emu/cm$^3$, exchange constants $A_{Py} = 1.3 \times 10^{-6}$ erg/cm and $A_{Co} = 2.0 \times 10^{-6}$ erg/cm, gyromagnetic ratios $\gamma_{Py}/2\pi = 2.96$ GHz/kOe and $\gamma_{Co}/2\pi = 3.02$ GHz/kOe.

The static and dynamic properties of these magnetic systems have been investigated by means of two micromagnetic codes: Object Oriented MicroMagnetic Framework (OOMMF) code[42] and DMM program[34,44]. The ground-state magnetization was determined by using OOMMF with 2D periodic boundary conditions; then this magnetic configuration was used as input to DMM. The DMM with implemented boundary conditions, a finite-difference micromagnetic approach first implemented for isolated ferromagnetic elements and extended to MCs composed by two ferromagnetic materials[34], is applied to study the spin dynamic properties in bi-component systems where the two ferromagnetic materials are separated by a non-magnetic spacer. Since our results do not focus on dissipation properties of collective modes, the dynamics is studied in the purely conservative regime, hence no Gilbert damping energy density contribution is included in the equations of motion. For our purposes, in the DMM two indexes are used: 1) an index $k$ to label micromagnetic cells, with $k = 1,2,... N$, where $N$ is the total number of micromagnetic cells in the primitive cell; 2) an index $j$ = Py, Co indicating the ferromagnetic material. The number of micromagnetic cells assigned to the *j*-th ferromagnetic material is $N_j$ such that $N_{Py} + N_{Co} = N$. For each micromagnetic cell the magnetization in reduced units takes the form $\mathbf{m}_k = \mathbf{M}_k / M_s(k)$ with $\mathbf{M}_k$ the magnetization in the *k*-th cell and $M_s(k)$ the saturation magnetization depending on the ferromagnetic material through the index *k*. Hence, in a polar reference frame the magnetization can be written in form:

$$\mathbf{m}_k = (\sin\theta_k \cos\phi_k, \sin\theta_k \sin\phi_k, \cos\theta_k) \qquad (1)$$

where $\phi_k$ ($\theta_k$) is the azimuthal (polar) angle of the magnetization; for the sake of simplicity the time dependence is omitted. The total energy density $E = \dfrac{\tilde{E}}{V}$, with $\tilde{E}$ the energy and *V* the volume of the system, respectively, depends on the polar and azimuthal angle in each micromagnetic cell, $\theta_k$ and $\phi_k$. The total energy $\tilde{E}$ is the magnetic Hamiltonian and the DMM was developed to study conservative systems corresponding to a purely precessional dynamics. In explicit form, for the systems under study, the energy density reads:

$$E = E_{\text{ext}} + E_{\text{exch}} + E_{\text{dmg}} \qquad (2)$$

with $E_{\text{ext}}$ the Zeeman, $E_{\text{exch}}$ the exchange, $E_{\text{dmg}}$ the demagnetizing, respectively. Specifically:

$$E = -M_S \mathbf{H} \cdot \sum_{k=1}^{N} \mathbf{m}_k + \sum_{k=1}^{N} \sum_{n \in \{\text{n.n.}\}} A_{\text{exch}}(k,n) \frac{1 - \mathbf{m}_k \cdot \mathbf{m}_n}{a_{kn}^2} + \frac{1}{2} M_S^2 \sum_{kl} \mathbf{m}_k \cdot \overrightarrow{\mathbf{N}} \mathbf{m}_l \qquad (3)$$

The first term of the Eq. (3) corresponds to the Zeeman energy density, where **H** indicates the external magnetic field. The second term of Eq. (3) is the exchange energy density expressed by means of two sums: the first sum runs over the *N* micromagnetic cells and is indexed by *k* whereas the second sum indexed by *n* ranges over the nearest neighbouring (n.n.) micromagnetic cells of the *k*th micromagnetic cell that can belong to a different ferromagnetic material. $A_{\text{exch}}$ is the exchange stiffness constant and is related to the ferromagnetic materials through the indexes *k* and *n*, respectively, while $a_{kn}$ denotes the distance between the centers of two adjacent micromagnetic cells of indexes $k$ and $n$, respectively. When the *k*th micromagnetic cell is on one of the edges (vertices) of the proper primitive cell, the interaction with the micromagnetic cells belonging to the correct nearest supercell (primitive cell) must be taken into account. The last term of Eq. (3) is the demagnetizing energy density where $\overrightarrow{N}$ is the demagnetizing tensor and expresses the interaction among micromagnetic cells within the primitive cell and belonging to different primitive cells. Note that, unlike the bi-component system studied[34] in S2$^{Co}$ the intermaterial exchange contribution is set equal to zero, because in the primitive cell the Co dot and the Py matrix are separated. Instead, in the S1, the exchange contribution at the interface between the two ferromagnetic materials is set equal to $A_{\text{exch}}^{Py-Co} = (A_{Py} + A_{Co})/2$ because Py matrix and Co dots are in contact. Note that in Eq. (2) the thermal contribution related to the thermal field is not included. Indeed, the studied dynamics is purely deterministic and not stochastic. Actually, the equations of motion within the DMM correspond to the deterministic Landau-Lifshitz equations and not to the stochastic Langevin or Fokker-Planck equations[45].

The dynamic magnetization $\delta\mathbf{m}(\mathbf{r})$ of each collective mode fulfils the generalized Bloch theorem depending on the Bloch wave vector **K** and on the two-dimensional lattice vector of the periodic system **R**. For each micromagnetic cell $\delta\mathbf{m}(\mathbf{r})$ is expressed in polar coordinates depending on the angular deviation from the equilibrium position of the azimuthal and polar angles $\delta\phi_k, \delta\theta_k$. In a compact form, the complex generalized Hermitian eigenvalue problem takes the form

$$\underline{\underline{A}}\mathbf{v} = \lambda \underline{\underline{B}}\mathbf{v}, \qquad (4)$$

where the eigenvalue $\lambda = \frac{1}{\omega}$ with $\omega$ the angular frequency of the given collective mode which is in turn described by the eigenvector $\mathbf{v} = (\delta\phi_k, \delta\theta_k)$. The Hermitian matrix $\underline{\underline{A}}$ depends on saturation magnetization of the two ferromagnetic materials and on the corresponding gyromagnetic ratios. The Hessian matrix $\underline{\underline{B}}$ is expressed in term of the second derivatives of the energy density with respect to the azimuthal and polar angular deviations $\delta\phi_k$ and $\delta\theta_k$ calculated at equilibrium. For further technical details of the DMM applied to several materials see Ref. 34.

The use of the DMM for calculating the spectrum of collective spin wave modes is preferred with respect to the Fourier analysis of OOMMF because it has several computational advantages. Among them, just to mention a few: a) the system under study does not need to be excited by any magnetic field pulse, b) the spin-wave modes frequencies and eigenvectors of any symmetry are determined by means a single calculation, c) the spatial profiles of the spin-wave modes are directly connected to the calculated eigenvectors allowing to accurately classify each collective excitation, d) the spectrum is computed directly in the frequency domain, e) the mode degeneracy is completely taken into account, f) the differential scattering cross-section associated to each spin-wave mode can be computed accurately starting from the corresponding eigenvectors. The size of the micromagnetic cells used in the static and dynamic simulations is $5 \times 5 \times 10$ nm along *x*, *y* and *z*, respectively.

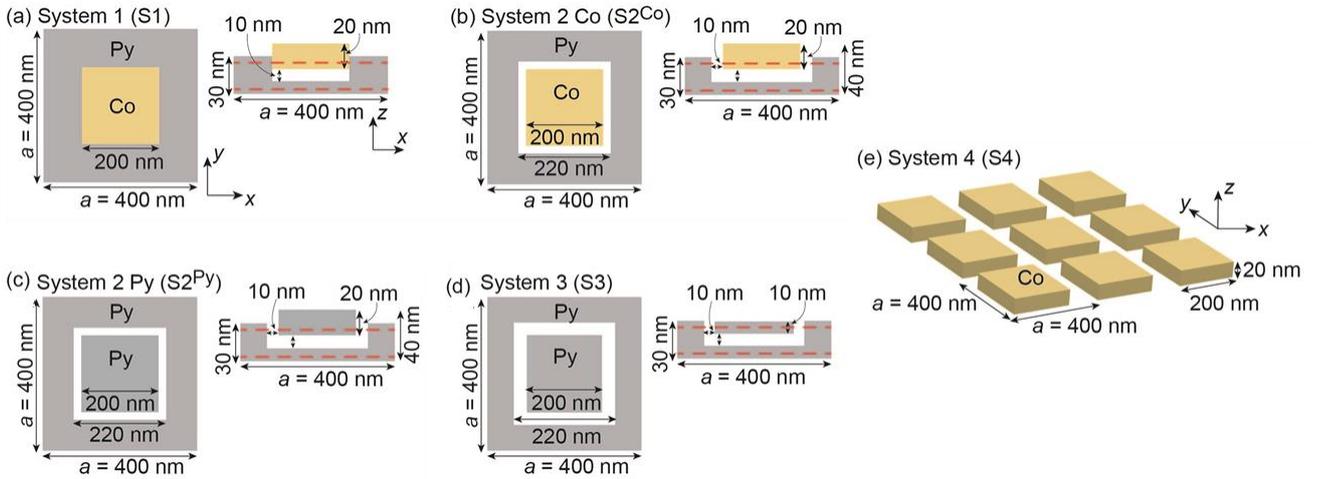

**Fig. 1.** (a) System 1: Top view of the primitive cell and its perpendicular cross-section in a bi-component MC consisting of Co square dots in square array partially immersed in Py. Non-magnetic spacer (white area) of 10 nm thick separates the bottom of Co dots from Py. (b) System 2-Co: similar to S1 but with full separation of Co dots (200 nm wide) from Py (10 nm of non-magnetic spacer from the bottom and lateral sides of Co). (c) System S2-Py: one component MC with geometry equals to S2$^{Co}$ but with Py dot. (d) System 3: MC created by square array of square grooves in Py film partially filled with Py dots. Dots are separated from the matrix by 10 nm thick non-magnetic spacer. (e) System 4: square array of square Co dots. Red-dashed lines in the perpendicular cross-sections point at the planes ($z = 5$ and 25 nm) used in Figs. 2(a), 2(b), 2(c), 2(d) and 2(e) to visualize the spatial profiles of SW modes.

In order to investigate the propagation properties of SWs in MCs, the systems have been studied in the DE geometry, i.e., with the external magnetic field (**H**) of magnitude fixed at 2000 Oe parallel to the *y*-axis and the Bloch wave vector (**k**) parallel to the *x*-axis.

**III. Spin wave excitations in MCs**

In 2D antidot lattices and bi-component MCs a full magnonic spectra is very rich with plenty of SW excitations[33]. As an example, the differential scattering cross-section computed at the center of the BZ is displayed in Fig. 2 for S1. It can be seen that there is a large number of spin wave modes resulting from the calculation. However, for the purposes of this study focused on the dispersion behavior in the first BZ only three modes belonging to the lowest frequency part of the spectrum, namely the ones exhibiting an appreciable differential scattering cross-section, have been selected in S1. Nevertheless, note that there are also other collective modes in the highest frequency part of the spectrum having a non-negligible differential scattering cross-section, but in higher BZs. The same conclusions on the differential scatting cross-section can be drawn also for the other systems. The dispersion relations shown in Fig. 3 are the ones measured in a typical BLS experiment[32,46].

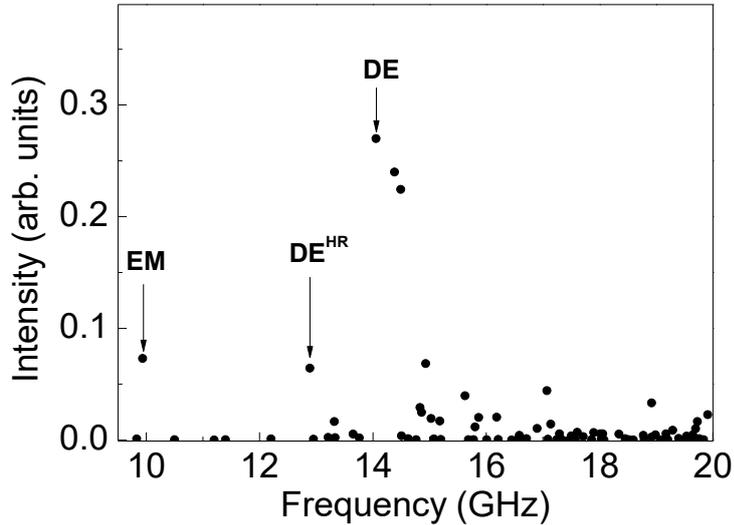

**Fig. 2.** Differential scattering cross-section calculated for S1 at the center of the BZ. The arrows label the modes with the highest scattering cross section in the center of the first BZ investigated in this paper.

The dispersion relations of SWs in S1 are shown in Fig. 3(a). We classify the collective modes by taking into account the region inside the primitive cell where they have the maximum amplitude. In this respect, we named them: 1) end mode of the dot ($EM_d$) (where the subscript "d" labels dot) mainly localized at the borders of the dots, 2) Damon-Eshbach-like mode in horizontal rows ($DE^{HR}$) where the superscript "HR" means horizontal rows and 3) Damon-Eshbach-like (DE) mode; they have frequencies 9.94, 12.89 and 14.06 GHz, respectively. The modes 2) and 3) are called Damon-Eshbach-like because they exhibit nodal planes parallel to the local static magnetization in the higher BZs and no nodal planes in the center of the BZ (see, Fig. 4(a)). This is in accordance to the classification of collective modes given for binary magnonic crystals[34]. In the center of the BZ, the $DE^{HR}$ and DE are the resonance modes called fundamental modes. The $DE^{HR}$ mode is localized in the horizontal rows containing the square dots (with amplitude concentrated mainly in Py), while the DE mode has the maximum amplitude in Co dots and non-negligible amplitude in the Py film. We point out that the end mode detected here has been previously found only in one component MCs[47,48].

The appearance of the end mode and the different SW amplitude distribution between Py and Co of DE and $DE^{HR}$ modes marks the difference between the S1 and the Co/Py bi-component MC investigated in Refs. [14,34]. We remark that these differences with respect to previous studied systems are mainly due to: a) the 10 nm thick non-magnetic spacer between Co dots and Py matrix placed at the bottom of the dots, and b) the dot shape (these effects will be discussed in the next paragraph). Next, we study the effect of a full separation of Co dots from Py matrix on magnonic spectra. In Fig. 3(b), the dispersion curves for $S2^{Co}$ are presented. By looking at Fig. 3(b) we can see the appearance of two new modes, i.e. the end mode of Py film ($EM_f$) at 11.9 GHz (where the subscript "f" labels film) localized at the border of Py film and the backward-like mode ($BA^{HR}$) at 13.86 GHz mainly concentrated in the horizontal rows. $BA^{HR}$ mode has nodal planes perpendicular to the local static magnetization (see, Fig. 4(b) for profiles of the modes). The frequency of the $BA^{HR}$ in $S2^{Co}$ is higher than the frequency of $DE^{HR}$. This might be attributed to the strong localization feature of the $BA^{HR}$ in the region filled by Co dots having higher values of the magnetic parameters. By comparing frequency at the center of the BZ passing from S1 to $S2^{Co}$, we observe a significant decrease of the $EM_d$ frequency from 9.94 GHz to 5.47 GHz and a slight increase of the DE ($DE^{HR}$) frequencies from 14.06 GHz (12.89 GHz) to 14.67 GHz (13.48 GHz). The presence of five dispersion curves in $S2^{Co}$ is attributed to the fact that the differential scattering cross-section is comparable for the five SW excitations at the BZ center.

In order to study the effect of the Py matrix on the SW excitation in Co dots, we calculate the dispersion curves of S4 [Fig. 1 (e)], array composed of square Co dots. By inspection of Fig. 3(e) we note that the frequency of the $EM_d$ in S4 (3.5 GHz) is about 6 GHz lower than in S1 and 2.5 GHz lower as compared to the corresponding one in $S2^{Co}$. Instead, the frequency of the DE mode (18.4 GHz) is 4 GHz higher than the one in S1 and 3.5 GHz higher than the one in $S2^{Co}$. Therefore, the effect of Py matrix is to lower the frequencies of the DE mode and to raise the frequencies of the $EM_d$. This behavior can be understood by taking into account the variation of the magnitude of the interdot dipolar dynamic coupling and of the static demagnetizing field passing from an array of dots (S4) to a MCs (S1 and $S2^{Co}$) composed by two ferromagnetic materials.

To study the effect of dot material and thickness in a Py matrix, we calculate the SW spectra of 2D MCs composed of Py dots in a Py matrix. It is important to underline that $S2^{Py}$ and S3 are neither a bi-component MC nor antidot lattices, but these structures preserve properties of both with the use of a single ferromagnetic material. The kind of modes found in $S2^{Py}$ and S3 is similar to the one found in $S2^{Co}$. In $S2^{Py}$, the $EM_d$ (8.92 GHz) is the lowest frequency mode as in $S2^{Co}$. In $S2^{Py}$ the $EM_f$ (10.46 GHz) has a dispersion curve similar to that of $EM_d$. The DE mode has a frequency of 12.8 GHz at the center of the BZ. The frequencies of $BA^{HR}$ mode (13.8 GHz) is lower than the ones of the $DE^{HR}$ mode (14.12 GHz). We observe that in $S2^{Py}$ the frequency sequence of DE, $DE^{HR}$ and $BA^{HR}$ modes is different with respect to the one in S1 and $S2^{Co}$ (see Figs. 3(a) and 3(b)). In particular, the DE mode frequencies are lower than the $DE^{HR}$ mode ones as in the case of 2D one component antidot lattices[47,48] (for further discussion see Section IV).

In order to understand the effect of the thickness of Py dots we compute the dispersion curves for S3 shown in Fig. 3(d). In S3, the $EM_f$ (8.92 GHz) is the lowest frequency mode of the spectrum. The $EM_d$ frequency at the center of the BZ (12.84 GHz) is larger than the one of the DE mode (12.36 GHz), however the corresponding dispersion curves have a similar behavior. This frequency inversion as compared to $S2^{Py}$ is not surprising because the total magnetic field experienced by the $EM_d$ is higher with respect to the field felt by the DE. The $DE^{HR}$ and $BA^{HR}$ have frequencies 14.68 and 14.12 GHz at the center of the BZ, respectively. Comparing the dispersion curves in $S2^{Co}$ and S3, we observe that the order of DE and $DE^{HR}$ frequency modes in S3 is interchanged with respect to the ones in $S2^{Co}$. Moreover, also the frequency order of the $EM_f$ and the $EM_d$ is interchanged with respect to the one in $S2^{Py}$ and $S2^{Co}$. This interchange can be attributed to the effect of the reduction of the dot thickness that induces a lowering of the total magnetic field in the Py film where the $EM_f$ is localized. The intensities of the differential scattering cross-section of the DE, $EM_d$, $EM_f$ and $BA^{HR}$ modes are comparable but are 40% lower than that of $DE^{HR}$.

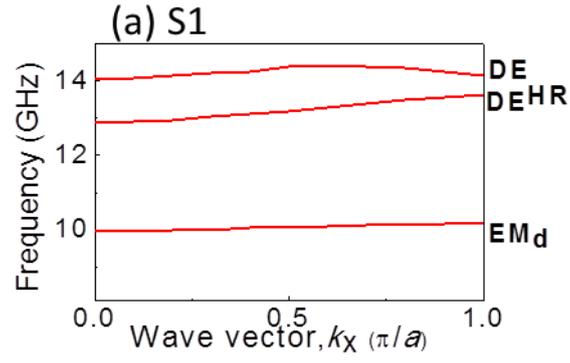
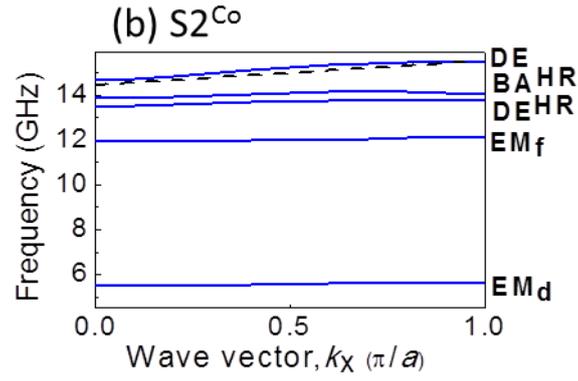
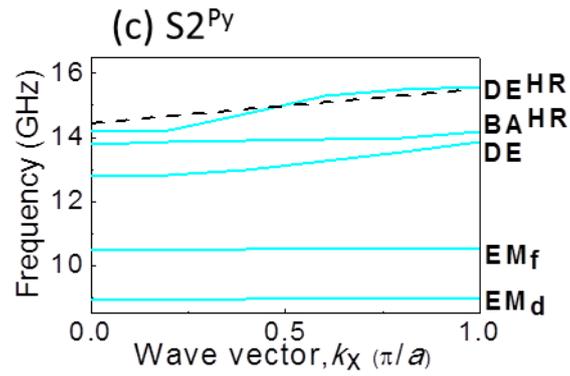
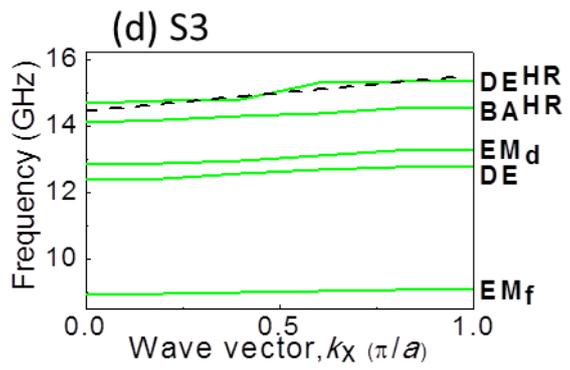
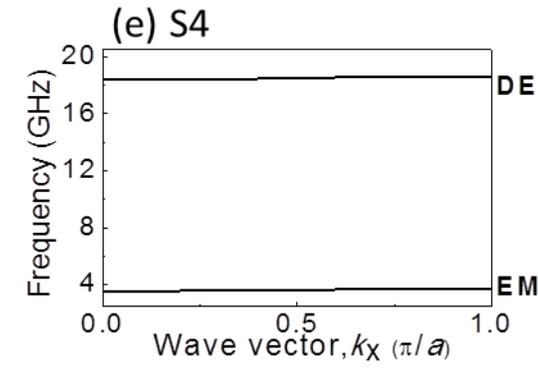

**Fig. 3.** Dispersion relation in the first BZ along the direction perpendicular to the external magnetic field. (a) Dispersion relation of the most relevant modes in S1: end mode of the dot (EM$_d$), Damon-Eshbach-like mode (DE) and DE-like mode in horizontal rows (DE$^{HR}$) are shown. (b) Dispersion relation in S2$^{Co}$. The additional dispersion relation of the end mode in Py film (EM$_f$) and backward-like volume SW (BA$^{HR}$) are shown. (c) Dispersion relation of the most relevant modes in S2$^{Py}$. (d) Dispersion relation in S3. (e) Dispersion relation in the array of Co dots (S4). The black dashed lines in Figs. 3(b), (c) and (d) mark dispersion relation of DE mode in homogeneous Py film of 10 nm thickness calculated according to Ref. 49.

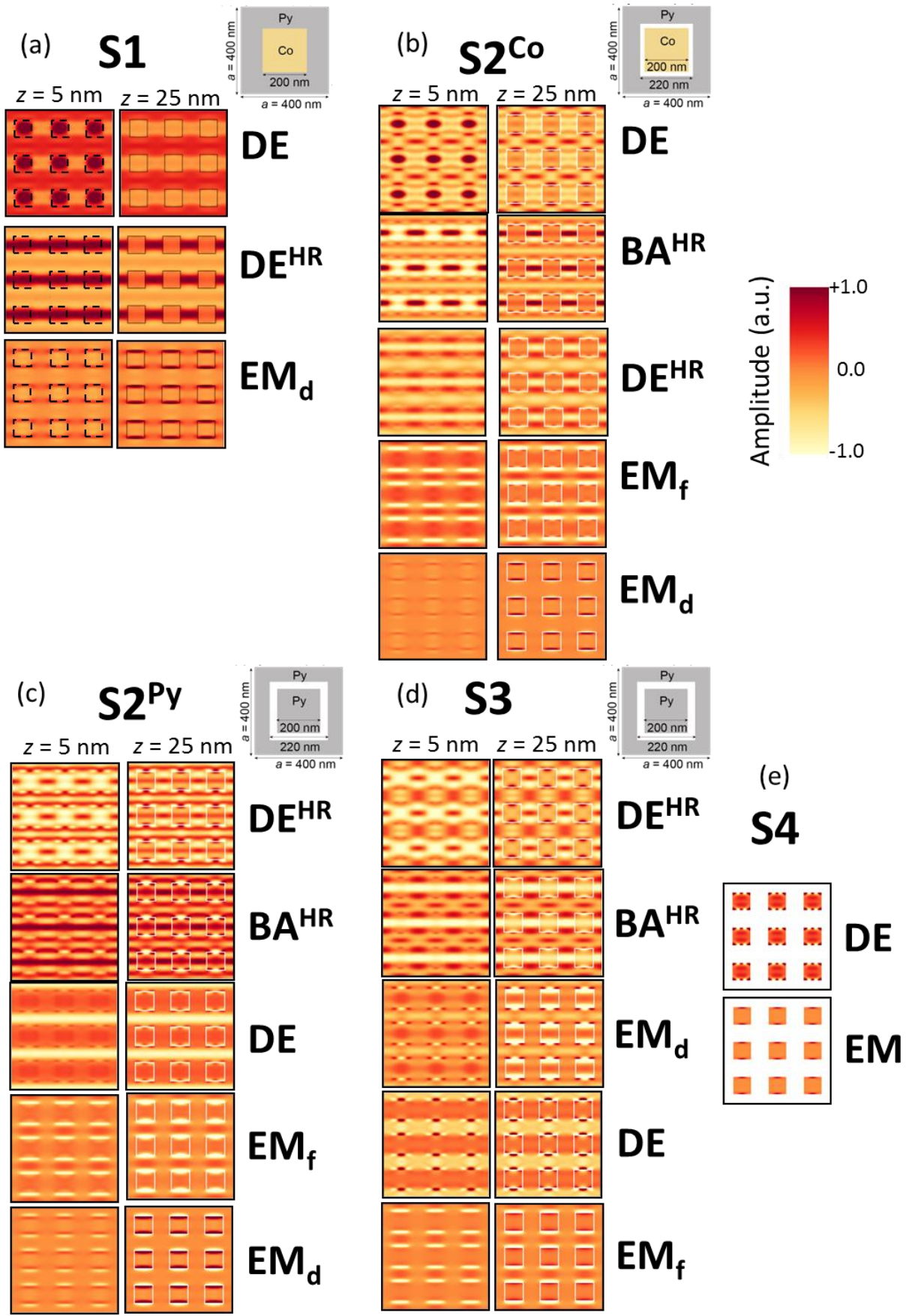

**Fig. 4.** Spatial profiles (real part of the out-of-plane component of the dynamic magnetization vector) for SWs with large differential scattering cross-section in the center of the Brillouin zone. The spatial profiles of SW modes from the bottom part of the Py film (in the plane $z = 5$ nm in left column) and in the plane crossing dots (for $z = 25$ nm in right column) are shown in 3×3 primitive cells, i.e., on the planes marked in Fig. 1 with red dashed lines. (a) Spatial profiles of EM, $DE^{HR}$ and DE modes in S1. (b) Spatial profiles of $EM_d$, $EM_f$, $DE^{HR}$, $BA^{HR}$ and DE modes in $S2^{Co}$. (c) Spatial profiles of $EM_d$, $EM_f$, DE, $BA^{HR}$ and $DE^{HR}$ modes in $S2^{Py}$. (d) Spatial profiles of $EM_f$, DE, $EM_d$, $BA^{HR}$ and $DE^{HR}$ modes in S3. (e) Spatial profiles of EM and DE modes in S4.

In Fig. 4 we show the spatial profiles of the real part of the out-of-plane component of the dynamic magnetization for the main modes at the center of the BZ of the systems studied. The spatial profiles are presented at planes $z = 5$ nm and $z = 25$ nm, left and right column of each panels respectively, along the cross-sections indicated in Fig. 1 with red-dashed lines. Looking at Fig. 4(a) we can see that the $EM_d$ is strongly localized at the border of Co dots and its amplitude decreases at $z = 5$ nm where only Py is present with respect to $z = 25$ nm. The presence of the Co dots in S1 induces a strong $DE^{HR}$ amplitude decrease inside the region containing the Co dots: indeed, for $z < 10$ nm the amplitude of the $DE^{HR}$ mode is uniform in the whole rows, while for 20 nm $< z <$ 30 nm its amplitude decreases in the Co dots region. By contrast, for the $EM_d$ the square Co dots induce an opposite behavior. The amplitude distribution of the DE mode takes contribution from both Co dots and Py matrix through its whole thickness. The DE is also the mode with largest differential scattering cross-section. Its intensity at $k_x = 0$ is three times larger than that of the $EM_d$ or the $DE^{HR}$ mode (see Fig. 2). Fig. 4(b) displays the spatial profiles of the characteristic SW modes of $S2^{Co}$. The presence of the non-magnetic spacer around the Co dots induces the appearance of the $EM_f$ that is strongly localized at the border of the Py matrix close to the non-magnetic spacer. The amplitude of this mode is almost uniform along the thickness, while that of the $EM_d$ decreases by decreasing $z$. The $DE^{HR}$, $BA^{HR}$ and DE modes have uniform amplitude in the region of Py matrix along the thickness. On the other hand, in the region filled by Co dots their amplitude strongly decreases for $z > 20$ nm.

In Fig. 4(c) we show the spatial profiles of the collective excitations in $S2^{Py}$. The amplitude variation of the $EM_d$, DE, $BA^{HR}$ and $DE^{HR}$ modes as a function of $z$ is the same as that in $S2^{Co}$. Moreover, in $S2^{Py}$ the amplitude of $EM_f$ decreases by decreasing $z$ following a trend similar to that of the $EM_d$. The amplitude of SW modes of S3 are illustrated in Fig. 4 (d). Similarly to what occurs in $S2^{Co}$ and $S2^{Py}$, the SW amplitude of the DE mode is almost homogeneous across the thickness of the whole structure and larger in the rows between dots. The $DE^{HR}$ and $BA^{HR}$ modes amplitude is almost uniform along $z$ in the Py matrix but decreases for $z > 20$ nm in the region filled by Py dots. In Fig. 4(e) are depicted the spatial profiles of collective modes in S4. In this system there is only Co along $z$ and the amplitudes of EM and DE mode are uniform along the thickness.

**IV. Total magnetic field analysis**

In order to understand the dispersion curves of the investigated structures, we calculate the in-plane components of the total (effective) magnetic field at different values of $z$. The total static magnetic field, which is the sum of the exchange field, the demagnetizing field and the Zeeman field, calculated for each micromagnetic cell by the OOMMF code, is averaged along the $x$ direction for different values of $z$ and $y$. The behavior of the total magnetic field is strictly related to the orientation of the static magnetization in the magnetic system. In Fig. 5 four regions along the thickness are taken into account: a) 0 nm $\leq z \leq$ 10 nm where only Py is present; b) 10 nm $< z \leq$ 20 nm where there are Py and non-magnetic spacer; c) 20 nm $< z \leq$ 30 nm where in S1 there are Py and Co, in $S2^{Co}$ there are Py, non-magnetic spacer and Co, in $S2^{Py}$ and S3 are present Py and non-magnetic spacer; d) 30 nm $< z \leq$ 40 nm where in S1 and $S2^{Co}$ there is Co, in $S2^{Py}$ is present Py.In particular, the presence of a well or a wall in the total magnetic field (see Fig. 5) is due to the saturation magnetization contrast present at interfaces between two different materials. Moreover, in MCs showing magnetization inhomogeneities across the thickness, the total magnetic field at interfaces between two materials, present for 10 nm $< z <$ 30 nm, influences also collective excitations in the homogeneous part of the structure (for 0 nm $< z <$ 10 nm).

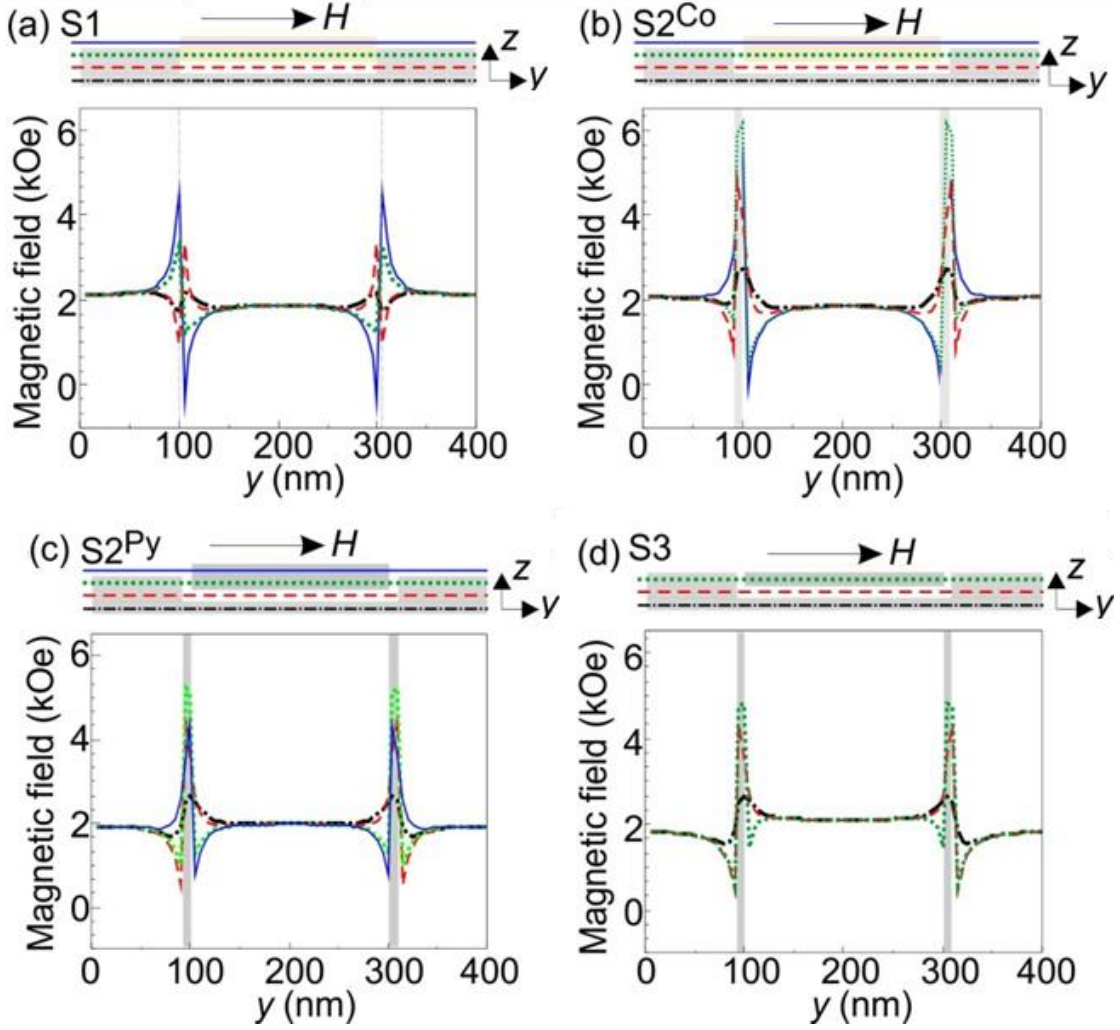

**Fig. 5.** The $y$ component of the total magnetic field calculated for (a) S1, (b) S2$^{Co}$, (c) S2$^{Py}$ and (d) S3 along the $y$-axis and averaged along $x$, for four different values of $z$: $z = 5$ nm (in full Py film, black dot-dashed line), $z = 15$ nm (crossing Py and spacer below the dots, red dashed line), $z = 25$ nm (crossing Py matrix and middle of dots, green dotted line) and $z = 35$ nm crossing Co dots (only in S1 and S2$^{Co}$, blue solid line). The gray vertical rectangles mark the non-magnetic spacer which separate the dot from the matrix. The insets on the top show a sketch of MCs with lines along which the total magnetic field is calculated.

The appearance of end modes in MCs is related to the presence of a strong inhomogeneity of the total field resulting in deep wells close to the border of the dots and the matrix. This feature of the total magnetic field in 2D bi-component MCs depends on two main factors: the shape of the dot and the contrast between the saturation magnetization of the different materials. In particular, the magnetization saturation contrast enhanced by the presence of the non-magnetic spacer leads to the formation of an inhomogeneous demagnetizing field and, as a consequence, to strong inhomogeneities of the total magnetic field at the border between two materials (Co/Py, Co/non-magnetic spacer and Py/non-magnetic spacer). Therefore, the presence of a thin non-magnetic spacer between two ferromagnetic materials not only influences significantly the SW spectra but can also be an end mode's creating factor. We underline that this important feature, namely the appearance of end modes, either as $EM_f$ or $EM_d$, does not depend on the dot shape or on the ferromagnetic material for MCs having geometric parameters in the range of the ones typical of the recently studied bi-component systems. Hence, this picture is different from the one occurring in bi-component systems[14,34] where a crucial rule to determine the appearance of end modes was played by a specific combination of the magnetization saturation contrast and the dot shape. As an example, in a bi-component MC composed of circular Co (Py) dots in direct contact with a Py (Co) matrix, the end mode is present when $|\Delta M_S| = |M_{S,Co} - M_{S,Py}| > 250$ emu/cm$^3$, but disappears when $|\Delta M_S| = 200$ emu/cm$^3$. Instead, if the bi-component system is composed of square Co (Py) dots in direct contact with Py (Co) matrix, an end mode is present when $|\Delta M_S| > 200$ emu/cm$^3$. Therefore, a thin non-magnetic spacer between the two ferromagnetic components of MCs not only influences significantly the SW spectra but also is an end mode's creating factor. We underline this important feature, namely the appearance of end modes, either as $EM_f$ or $EM_d$, does not depend on the dot shape or on the ferromagnetic material. In the following, we discuss the shape of the total magnetic field in 2D bi-component MCs introduced by non-magnetic spacers around dots and its relation to the end modes. Fig. 5(a) shows the total magnetic field calculated for S1 vs. $y$ for different values of $z$. Two deep wells are present inside the region of the Co dot above the Py matrix corresponding to $z > 30$ nm. The two wells are still present for 20 nm $< z <$ 30 nm, although with decreasing depth. The two wells disappear for $z < 20$ nm, however the walls appear in this range. For this reason the $EM_d$ is strongly localized in the well of the total magnetic field at the border of Co dot for $z > 20$ nm and disappears in the homogeneous part of the system where there is the Py matrix ($z < 20$ nm) [see Fig. 4(a)].

In Fig. 5(b) it is displayed the total magnetic field calculated for S2$^{Co}$ as a function of $y$ for different values of $z$. It can be seen that the positions of the minima of the total magnetic field depend on $z$. In particular, the total magnetic field has its minimum value in the Py region for $z < 20$ nm, while in the Co region for $z > 20$ nm. These two wells close to the border between Py and the non-magnetic spacer and the non-magnetic spacer and Co give rise to the two localized modes $EM_f$ and $EM_d$, respectively. Thus, the presence of these two end modes is strictly related to the non-magnetic material that surrounds the Co dots responsible for the appearance of the two minima in the total magnetic field.

Comparing the profiles of the total field at $z = 15$ nm and $z = 25$ nm (red dashed and green dotted line in Figs. 5(a) and 5(b)), an increase of the depth of the magnetic wells can be noted in S2$^{Co}$ with respect to the one in S1. This explains the decrease of the frequency of the $EM_d$ in S2$^{Co}$ as compared to the one in S1. Moreover, the wells of the total field corresponding to the region filled by the Py matrix close to the non-magnetic spacer at $z = 15$ nm, although less deep than the ones in the Co dot, are deep enough to permit localization of the $EM_f$.

By looking at Fig. 5(a) it is also possible to understand that the variation of the total magnetic field due to the non-magnetic spacer induces a change of DE$^{HR}$ and DE mode profiles as a function of $z$. We observe that the uniform amplitude of DE$^{HR}$ in the horizontal rows (see Fig. 4(a)) is due to the trend of the total magnetic field. Indeed, by looking at Fig. 5(a) (black dot-dashed line), we note that the total magnetic field does not present significant inhomogeneities along the $y$ direction at $z = 5$ nm. Instead, at $z = 25$ nm the DE$^{HR}$ mode is localized only in the Py region (see Fig. 4(a)) and its amplitude vanishes inside the Co dot. On closer inspection of the corresponding total magnetic field [Fig. 5(a) green dotted line] we note the presence of a high wall at the border between Py and Co that prevents the spreading of DE$^{HR}$ inside the Co dot. The DE mode has higher frequency than DE$^{HR}$ and its amplitude spreads also in Co dot for $z > 20$ nm. In S2$^{Co}$, there is an increase of the total magnetic field inhomogeneity as compared to S1 for each value of $z$, apart from $z > 30$ nm where there is a small reduction [see Figs. 5(a) and (b)]. This results in an increase of the frequencies of the DE and DE$^{HR}$ modes.

In Fig. 5(c) we plot the total magnetic field for S2$^{Py}$ in order to investigate the effect of the change of the material filling the dots. There are two minima of the total magnetic field: the absolute minimum is located in the Py matrix for 10 nm $< z <$ 20 nm and the other minimun is placed in the Py dot for 30 nm $< z <$ 40 nm. In corrispondence of the above mentioned minima, also in S2$^{Py}$ there is the appearance of the $EM_d$ and of the $EM_f$, respectively. By looking at Figs. 5(b), (c) and (d), we can note a qualitative similarity of the behavior of the total magnetic field as a function of *y* in S2$^{Co}$, S2$^{Py}$ and S3, respectively. In Fig. 5(d), where the Py dot thickness is 10 nm, the magnetic field well in the dot is less deep than the one in S2$^{Py}$, while in the Py matrix it has a significant minimum (green dotted line, *z* = 25 nm). This explains the interchange of the frequencies of the EM$_f$ and EM$_d$ modes found in S3 with respect to the ones in S2$^{Co}$ and S2$^{Py}$. Detailed inspection of the total magnetic field profiles shown in Fig. 5 (b), (c) and (d) allows to notice also the relative change of the magnetic field values among S2$^{Co}$, S2$^{Py}$ and S3 in the channels parallel to the *x*-axis containing dots [i.e., area of the DE$^{HR}$ mode, for 100 nm < *y* < 300 nm in Fig. 5] and lying between the dots [i.e., area of the DE mode for 0 nm < *y* < 90 nm and 310 nm < *y* < 400 nm]. In the middle part of these areas the average value of the total magnetic field is almost constant across the full thickness. In S2$^{Co}$ the values of the field are 2.06 and 1.85 kOe in center of the areas of DE and DE$^{HR}$ mode, respectively, while in S3 the respective values are 1.82 and 2.1 kOe. This behavior of the field can explain the frequency exchange of the DE and DE$^{HR}$ modes between S2$^{Co}$, S2$^{Py}$ and S3 in Fig. 3 (b), (c) and (d), respectively.

## V. Features of the dispersion relation

In order to fully understand the effect of different position and size of the non-magnetic spacer on the propagation of SWs, we compute the group velocity and the band width for the most relevant modes. The group velocity is important e.g. in the transmission measurements with the use of coplanar waveguide transducers, where SW with low wavenumber are usually excited[7]. A wide band width is important in order to accommodate incoming and transmitted signal; moreover, it can be used as an indicator of the interaction strength in the MC. The group velocity ($v_g$) in the DE geometry has been calculated for selected modes close to the center of the BZ, as:

$$v_g = 2\pi \frac{\Delta v}{\Delta k_x}, \qquad (5)$$

where $\Delta v$ is the change of the SW frequency due to the change of the wavevector along the *x*-axis, $\Delta k_x$ (in calculations we set $\Delta k_x = 0.05\ \pi/a$). The band width for selected mode has been calculated as a change of its frequency between BZ center and BZ border, $\Delta v_{bw} = |v(k_x = \pi/a) - v(k_x = 0)|$. The group velocity and the band width of the investigated SW excitations (EM$_d$, EM$_f$, DE and DE$^{HR}$) are calculated and collected in Table I.

**Table I.** Group velocity $v_g$ in the BZ center and band width for EM$_d$, EM$_f$, DE and DE$^{HR}$ modes in the MCs investigated in the paper. The two largest group velocities and band widths are emphasized in bold.

|  | S1 | | S2$^{Co}$ | | S2$^{Py}$ | | S3 | | S4 | |
| --- | --- | --- | --- | --- | --- | --- | --- | --- | --- | --- |
|  | $v_g$ [m/s] | Band width [MHz] | $v_g$ [m/s] | Band width [MHz] | $v_g$ [m/s] | Band width [MHz] | $v_g$ [m/s] | Band width [MHz] | $v_g$ [m/s] | Band width [MHz] |
| **EM$_d$** | 64 | 162 | 48 | 154 | 0 | 21 | 48 | 446 | 40 | 154 |
| **DE$^{HR}$** | 144 | 750 | **368** | 272 | 160 | **1378** | 160 | 668 | - | - |
| **DE** | 256 | 355 | **522** | 810 | 256 | **1097** | 152 | 410 | 68 | 203 |
| **EM$_f$** | - | - | 80 | 226 | 48 | 49 | 48 | 173 | - | - |

By looking at Tab. 1 we can see that for vanishing wave vector, the DE and $DE^{HR}$ modes in $S2^{Co}$ exhibit the largest group velocities. These larger values of $v_g$ can be attributed to a combination of higher contrast between Co and non-magnetic spacer and Py and non-magnetic spacer and to a higher Co gyromagnetic ratio. This is an interesting result as $S2^{Co}$ can be regarded as the most disruptive structure with respect to a homogeneous thin film. The $DE^{HR}$ modes in S1, $S2^{Py}$ and S3 have similar group velocities, while the DE mode of S3 has a group velocity smaller than the ones of the DE modes in S1 and $S2^{Py}$. The decrease of the group velocity in S3 can be due to the thickness reduction of the Py dots. These group velocities can be compared to that of the DE magnetostatic SW in homogeneous Py film of 10 nm thickness calculated according to Eq. (5). In this special case the latter turns out to be 880 m/s, a value larger than the ones of the systems studied as expected. The dispersion relation of the DE magnetostatic SW is superimposed in Figs. 3(b), (c) and (d) with black dashed line. We can see that it matches very well with the DE mode in $S2^{Co}$ and the $DE^{HR}$ modes in $S2^{Py}$ and S3. This shows that the DE and $DE^{HR}$ modes, in $S2^{Co}$, $S2^{Py}$ and S3 respectively, propagate in a way similar to that of the DE magnetostatic SW in homogeneous Py film and they travel mainly in the lower part of the structure where the dots influence on the internal field is smallest, nevertheless it changes the group velocity and band width.

Comparing the group velocities of DE and $DE^{HR}$ modes of S1, $S2^{Co}$, $S2^{Py}$, S3 and S4 with the one of the DE magnetostatic SW mode in homogeneous Py film, it can be noted that the presence of two different magnetic materials and a non-magnetic spacer reduces the speed of propagation in the BZ center. This is probably due to the presence of different magnetic material and non-magnetic spacer that induce the SW confinement in particular regions of the primitive cell.

The DE and $DE^{HR}$ mode of $S2^{Py}$ have the largest band width. It is interesting to note that also the end modes with higher frequency, $EM_f$ and $EM_d$ in $S2^{Co}$ and S3 have a band width comparable to that of the propagative $DE^{HR}$ and DE modes. This means that also the localized modes can propagate in this kind of MCs and their properties can be exploited for transmitting signal.

**V. Conclusions**

Detailed theoretical investigations of the spin wave spectra in two-dimensional bi-component MCs with the DMM, in order to identify the influence of a non-magnetic spacer on the magnonic band structure have been performed. Square arrays of square grooves in thin Py film filled (or partially filled) with Co or Py square dots have been studied. The conclusions drawn for these kind of MCs can be generalized to other kind of 2D lattices and of different dot shapes in the nanometric range. The non-magnetic spacer breaks exchange interactions between the magnetic materials of the matrix and the dot. However, most importantly, this non-magnetic spacer strongly modifies the total magnetic field, especially also at the dot edges. Due to these changes of the magnetic field, two types of end modes appear in the same structure. These are the end mode localized in the dot and that localized in the matrix. Their frequencies strongly depend on the magnetization of the matrix and of the dot material. Moreover, we have shown that, by employing a single material (Py in our case), it is possible to design a MC preserving the main properties of bi-component MCs and magnonic antidot lattices.

We have also shown that the introduction of a non-magnetic spacer and the change of the magnetic dot material allow to tailor in different ways the SW spectra in MCs. This includes even the interchange of the SW frequency order. This property can be further exploited for modeling the magnonic band structure and magnonic band gaps towards the properties desired for practical applications. Moreover, the non-magnetic spacer breaks the exchange interaction at the border between the two ferromagnetic materials and allows the fabrication of structures where magnetization reversal of the dots can take place at magnetic field values different from those causing magnetization reversal in the matrix (due to different shape or crystalline magnetic anisotropy). Here, there are more possibilities than in 1D re-programmable structures[50,51], because the anisotropy axis (and the magnetization) of the dots can be in an oblique direction with respect to the magnetization of the matrix.

The results of this study are interesting also for the investigation of the dynamical properties of bi-component MCs composed of hard and soft ferromagnetic materials, where stray magnetic field originating from the dots (made of hard ferromagnetic material) influences formation of the domain pattern[52] but SW dynamics has not been investigated so far in such structures.

## Acknowledgments

The research leading to these results has received funding from Polish National Science Centre project DEC-2-12/07/E/ST3/00538, from the European Union's Horizon 2020 research and innovation programme under the Marie Skłodowska-Curie grant agreement No 644348 (MagIC), and from MIUR-PRIN 2010-11 Project2010ECA8P3 "DyNanoMag".